\documentclass[12pt]{article}
\usepackage[utf8]{inputenc}
\usepackage{hyperref}
\usepackage{amsmath}
\usepackage{amssymb}
\usepackage{amsfonts}
\usepackage{bbm}
\usepackage{cite}
\usepackage{pifont}
\usepackage{mathtools}
\usepackage{esint}

\emergencystretch=\maxdimen
\title{Semiclassical Propagation and the Dynamics of Configuration Space}
\author{V.S. Morales-Salgado}
\begin{document}

\maketitle

\begin{abstract}
    This work explores the quantum propagator $K(x,t)$ as a solution of the system's dynamical equation.  
    We develop a generalized propagator framework in which the propagator is written in the form $K=\exp\!\left(R+\frac{i}{\hbar}S\right)$, where $S$ governs the semiclassical phase structure and \(R\) governs the amplitude transport and weighting of configurations.
    Starting from nonrelativistic quantum mechanics, the classical Hamilton--Jacobi equation emerges in the semiclassical limit, while $R$ reduces to the logarithm of the Van Vleck determinant.
    The formalism is then extended to relativistic field theory and minisuperspace quantum cosmology using functional methods and generalized Hamiltonian constraints.
    Then the resulting semiclassical equations recover the corresponding classical dynamics along characteristic flows.
    In minisuperspace models, the formalism yields coupled equations for geometric and matter sectors analogous to semiclassical Wheeler--DeWitt systems.
    In semiclassical gravitational settings, the gravitational contribution may admit an entropy-like interpretation, consistent with thermodynamic weighting factors appearing in Euclidean gravitational path integrals.
    
    The resulting framework suggests a unified semiclassical description in which propagator phases encode dynamical evolution while amplitude functionals encode transport and configuration weighting.
    This perspective motivates viewing mechanics as a relation between the differential structure of spacetime and the differential structure of the space of dynamical configurations.
\end{abstract}

\section{Introduction}
The propagator in quantum mechanics encodes the evolution of a system \cite{uv55,fh65}.
In particular, it is a solution of the system's dynamical equation \cite{uv55,fh65}.
On the other hand, for certain systems, solutions of the Schr\"odinger equation can be proposed to be proportional to an exponential of the classical action \cite{pm99}.
When valid, this assumption is referred to as the eikonal \cite{h23} or, in more general settings, the WKB approximation \cite{w26,k26,b26}.
For the propagator, this form appears in systems such as the free particle \cite{fh65}, the harmonic oscillator \cite{c98} along with its forced version \cite{h85}, and others \cite{dl84}.

Although this approach has been explored in the so-called de Broglie-Bohm pilot wave theory for the wave function, where it includes a deterministic interpretation of quantum mechanics \cite{b52,d22}, here we leave such interpretations aside.
Instead, we focus on the way the proposed form of the propagator helps us understand quantum dynamics.
The functional form of the propagator is independent of the interpretation of quantum theories, and we adhere to the usual postulates of quantum mechanics \cite{l18,h13}.
In particular, this work deals with the time evolution postulate by generalizing the Schr\"odinger equation \cite{cdl20}.

In this article, we revise propagators of the form $K\propto\exp(\mathrm{i}S/\hbar)$ and investigate how this approach can be generalized to dynamical equations of fields.
A key element of our investigation is the extra term $R$ in the exponent beyond the classical action $S$,
that is, propagators of the form:
\begin{equation}\label{ansatz}
    K = \exp\left(R+\frac{\mathrm{i}}{\hbar}S\right) .
\end{equation}

In this work, we primarily view $R$ as a \emph{transport function} governing the amplitude structure and the weighting of configurations as a function of spacetime coordinates.
In the semiclassical limit, $R$ reduces to the logarithm of the Van Vleck determinant (up to a constant).
In field-theoretic and gravitational contexts, $R$ may additionally admit a thermodynamic or entropy-like interpretation, as discussed in later sections.
These roles are compatible since the same mathematical object can describe quantum transport in one regime and thermodynamic weighting in another.

In the non-relativistic case, the simple condition of the proportionality constant depending only on time is straightforwardly connected to potentials quadratic in position.
However, more general cases require $R$ to be a function of both time and position, which we analyze in detail.

After establishing the non-relativistic setting, we propose an extension to field equations.
The key idea is to exploit the asymmetry between time and space in the Schr\"odinger equation and translate it into a distinction between spacetime coordinates and mechanical properties of fields.
As a result, several instances of field equations are produced, culminating in the case of a scalar field in a Robertson-Walker metric as a system whose dynamics is governed by a Hamiltonian constraint.

The approach presented here differs from standard semiclassical gravity and Wheeler--DeWitt WKB expansions in several respects.
In conventional WKB quantum cosmology, one expands the wave functional as $\Psi = \exp(i S_{\mathrm{cl}}/\hbar + \cdots)$ and obtains the Hamilton--Jacobi equation for $S_{\mathrm{cl}}$ plus a transport equation for the prefactor. Here we introduce an \emph{additive} exponent $R$ that is not restricted to be purely time-dependent; it can absorb both the usual WKB prefactor and additional complex contributions.
This allows us to treat the Hamiltonian constraint in a parallel semiclassical description: the classical constraint $H=0$ emerges only in the $\hbar\to0$ limit, while $R$ encodes quantum spreading and, in constrained systems, entropy-like terms.

Our formalism is closest in spirit to the \emph{quantum Hamilton--Jacobi} approach (e.g.,~ \cite{aks74}) and to \emph{decoherence} interpretations of the emergence of classicality.
However, we do not rely on a preferred time foliation.
Unlike Euclidean quantum gravity, where the entropy is obtained from the imaginary part of the Euclidean action, our complex-action split introduces an imaginary part directly in the Lorentzian Hamilton--Jacobi functional.
This parallels Jacobson's thermodynamic derivation of the Einstein equation \cite{j95}, but we work at the level of the propagator (generalized to a partition functional) rather than the Clausius relation.

We therefore position this work as an exploratory framework that generalizes the WKB/Van Vleck propagator to constrained systems, with a thermodynamic interpretation suggested by the structure of the complex action.
To situate our work, we briefly compare it with several well-established approaches.

In canonical quantum gravity, the Wheeler--DeWitt equation $\mathcal{H}\Psi = 0$ is often solved in the WKB approximation by writing $\Psi = \exp(i S_{\text{cl}}/\hbar + \dots)$. 
The leading order gives the Hamilton--Jacobi equation for $S_{\text{cl}}$, and the next order yields a transport equation for the prefactor. 
Our ansatz $K = \exp(R + iS/\hbar)$ generalises this by allowing $R$ to be a function of both time and the configuration variables, not merely a global prefactor. 
This flexibility becomes essential when treating the Hamiltonian constraint as emerging only in the $\hbar\to0$ limit.

The de Broglie-Bohm theory writes the wave function as $\psi = \sqrt{\rho} e^{iS/\hbar}$, leading to a Hamilton--Jacobi equation with an extra quantum potential $-\frac{\hbar^2}{2m}\frac{\nabla^2\sqrt{\rho}}{\sqrt{\rho}}$. 
Our $R$ plays a role analogous to $\frac12 \ln\rho$, but we do not impose a deterministic interpretation. 
Instead, we view $R$ as a transport function that can also absorb imaginary parts of $S$, extending the quantum Hamilton--Jacobi formalism (e.g.,~ \cite{aks74}) to constrained systems.

Jacobson's thermodynamic derivation of the Einstein equation \cite{j95} suggests that gravity may be an emergent phenomenon rooted in entropy. 
Our complex-action split $S = S_m + iS_g$ directly introduces an entropy-like term $\exp(-S_g/\hbar)$ in the propagator. 
This parallels the Euclidean action method where the imaginary part of the action gives the entropy, but here it appears in a Lorentzian framework. 
The closure ansatz is reminiscent of conditions arising from entropy extremisation.

In the Born-Oppenheimer approach to quantum cosmology, the heavy gravitational degrees of freedom are treated adiabatically, leading to a time-dependent Schr\"odinger equation for matter. 
Our separation $S = S_a(a,t) + S_\phi(\phi,t) + iS_g(a,t)$ follows a similar spirit: the imaginary part $S_g$ is associated with the gravitational sector and does not mix with matter. 
The resulting equations recover the classical Friedmann and Klein-Gordon equations without assuming a preferred time foliation.

Decoherence approaches explain the emergence of classicality by tracing out environmental degrees of freedom. 
Our propagator $K$ can be interpreted as a partition functional, and the factor $\exp(R/\hbar)$ weights different field configurations. 
In this sense, $R$ encodes both quantum spreading and a thermal-like factor, which could be linked to the influence functional in open quantum systems. 
A detailed comparison with stochastic gravity is left for future work.

Our framework does not replace any of the above methods; rather, it offers a formal correspondence (the exponential propagator ansatz) that accommodates many of their features. 
The key novelty is the treatment of $R$ as a dynamical field (or functional) that can carry both quantum amplitude information and thermodynamic entropy, while the classical equations emerge from the $\hbar\to0$ limit of the Hamilton--Jacobi sector.

A central theme of this work is that the dynamical equations may be interpreted as relations between two differential structures: the differential structure of spacetime coordinates and the differential structure of the space of mechanical properties (positions, fields, or geometric variables). 
In this perspective, the propagator does not merely describe trajectories in spacetime, but rather the propagation of constraints across a configuration space weighted by the semiclassical action.

The paper is organized as follows:
Section \ref{SchProp} reviews the case of propagators satisfying the Schr\"odinger equation, first with a time-only prefactor (which restricts to quadratic potentials), then with a space-and-time dependent $R$.
Section \ref{FldProp} extends the approach to field equations, treating fields as mechanical properties distributed over spacetime.
Section \ref{GRProp} applies the formalism to systems with a Hamiltonian constraint, specifically a scalar field in a Robertson-Walker metric and a matter field coupled to a conformal factor, where the role of $R$ becomes particularly transparent.
Section 5 discusses the conceptual implications of the formalism and the role of the transport functional.
Finally, Section 6 contains the conclusions and outlook.

\section{Schr\"odinger propagators}\label{SchProp}
In non-relativistic quantum mechanics, the propagator $K(x,t)$ satisfies the Schr\"odinger equation:
\begin{equation}\label{Sch}
    \mathrm{i}\hbar\partial_t K = H(K) ,
\end{equation}
where the Hamiltonian $H$ is an operator acting on $K$.
Although the propagator depends on both the initial point $(x_0,t_0)$ and the final point $(x,t)$, for simplicity we fix $(x_0,t_0)$ and omit it from now on, writing $K(x,t)$.

The Hamiltonian operator takes the standard form:
\begin{equation}\label{ClssH}
    H(K) = -\frac{\hbar^2}{2m} \partial_x^2 K + V K ,
\end{equation}
where $V=V(x,t)$ is the potential. This follows from the classical Hamiltonian $H = p^2/(2m) + V$ via the prescription of canonical quantization: $p \to -\mathrm{i}\hbar\partial_x$.

Our goal is to study propagators of the form (\ref{ansatz}), where $S$ is Hamilton's principal function (satisfying the Hamilton--Jacobi equation in the classical limit) and $R$ is an auxiliary function that we interpret as a \emph{measure of quantumness}.
In standard WKB theory, a purely time-dependent $R$ accounts for the semiclassical prefactor \cite{v28}.
Here we allow $R$ to depend also on position, which will enable us to capture quantum effects beyond the leading semiclassical approximation.
The function $R$ controls the spreading or focusing of the quantum wave packet relative to the classical trajectory.

Inserting (\ref{ansatz}) into (\ref{Sch}) and separating powers of $\hbar$ yields, to leading order, the Hamilton--Jacobi equation for $S$:
\begin{equation}\label{HJe}
    \partial_t S + \frac{1}{2m}(\partial_x S)^2 + V = 0 .
\end{equation}
The next order gives a constraint linking $R$ and $S$.

Before proceeding, we clarify the logical status of the equations that follow. The ansatz (\ref{ansatz}) is an exact substitution into the Schr\"odinger equation (\ref{Sch}).
No approximation is made at this stage.
The resulting equations are therefore exact consequences of the ansatz, not yet semiclassical approximations.

The semiclassical interpretation enters when we identify $S$ as Hamilton's principal function, which in the limit $\hbar\to 0$ satisfies the classical Hamilton--Jacobi equation (\ref{HJe}) independently of $R$; expand in powers of $\hbar$ and keep only leading terms, which yields the usual WKB approximation; and take the limit $\hbar\to 0$ explicitly to recover classical dynamics.
Thus, the ansatz is a formal device; the only approximation is the truncation of an $\hbar$-expansion or the classical limit.

We now explore two increasingly general families of solutions.

\subsection{Simple motivating examples}
The simplest case assumes that $R$ depends only on time.
This means the prefactor $\exp(R)$ is a global time-dependent normalization. Inserting (\ref{ansatz}) into (\ref{Sch}) and using (\ref{HJe}) yields a single remaining equation:
\begin{equation}\label{sch1s}
 \partial^2_x S + 2m\partial_t R = 0 .
\end{equation}
Physically, this equation relates the spatial curvature of the action $S$ (i.e.,~ how the classical momentum varies in space) to the time derivative of the quantumness function $R$.

Equation (\ref{sch1s}) can be integrated directly. Integrating twice with respect to $x$ gives:
\begin{equation}\label{St}
 S(x,t) = f_0(t) + f_1(t)x - m\partial_t Rx^2 ,
\end{equation}
where $f_0(t)$ and $f_1(t)$ are integration functions. Notice that $S$ is at most \emph{quadratic in $x$}. Substituting this form back into the Hamilton-Jacobi equation (\ref{HJe}) forces the potential to also be quadratic:
\begin{equation}\label{Vsqr}
    V(x,t) = g_2(t) x^2 + g_1(t) x + g_0(t) \,. 
\end{equation}
Hence, propagators of the form (\ref{ansatz}) with $R=R(t)$ correspond to quadratic potentials. This includes the free particle ($g_2=g_1=g_0=0$), the harmonic oscillator ($g_2 = m\omega^2/2$, $g_1=g_0=0$), and the driven harmonic oscillator ($g_2$ constant, $g_1=0$, $g_0$ arbitrary).

For each such potential, the functions $R$, $f_1$, $f_0$ are determined by a system of ordinary differential equations derived from (\ref{HJe}) and (\ref{St}):
\begin{align}
    m\,\partial^2_t R &= 2m(\partial_t R)^2 + g_2 , \\
    \partial_t f_1 + g_1 &= 2(\partial_t R) f_1  , \\
    2m(\partial_t f_0 + g_0) + f_1^2 &= 0 . 
\end{align}
Solving these equations for the cases mentioned above yields:
\begin{itemize}
\item Free particle ($g_2=g_1=g_0=0$):
\begin{equation}
    R = -\frac{1}{2}\ln(t) + \text{const} , \qquad S = \frac{m(x-x_0)^2}{2t}.
\end{equation}
The logarithmic decay of $R$ produces the familiar $1/\sqrt{t}$ prefactor.

\item Harmonic oscillator ($g_2 = m\omega^2/2$, $g_1=g_0=0$):
{\small
\begin{equation}
    R = -\frac{1}{2}\ln\bigl[\sin(\omega t)\bigr] + \text{const}, \qquad
    S = \frac{m\omega}{2}\left[(x_0^2+x^2)\cot(\omega t) - 2x_0 x \csc(\omega t)\right].
\end{equation}}
Here $R$ diverges at $t = n\pi/\omega$, marking the caustics where the propagator refocuses.

\item Driven harmonic oscillator ($g_2 = m\omega^2/2$, $g_1=0$, $g_0(t)$ arbitrary):
\begin{align*}
    \partial_t R &= \frac{\omega}{2} \tan(\omega t + c_0),\\
    f_1 &= c_1 \sec(\omega t + c_0),\\
    \partial_t f_0 &= \frac{m}{2}f_1^2 + g_0(t).
\end{align*}
The function $g_0(t)$ encodes the driving force.
\end{itemize}

Note that for quadratic potentials where $R=R(t)$, the prefactor $\exp(R)$ coincides with the inverse square root of the Van Vleck determinant:
\begin{equation}
    \exp(R) = \left(\det\left(-\frac{\partial^2 S}{\partial x \partial x_0}\right)\right)^{-1/2} = \frac{1}{\sqrt{D_{VV}}}.    
\end{equation}
Indeed, for the free particle, $R = -\frac{1}{2}\ln t$ gives $\exp(R) = 1/\sqrt{t}$, while $D_{VV} = \partial_x\partial_{x_0}S = m/t$.
For the harmonic oscillator, $\exp(R) = 1/\sqrt{\sin(\omega t)}$ matches the Van Vleck determinant up to a constant phase.
This identification supports our interpretation of $R$ as a transport function controlling amplitude spreading.

\subsection{A more general case}
The preceding analysis shows that restricting $R$ to be time-dependent leads to a severe limitation: the potential must be at most quadratic in $x$; most physically interesting systems fall outside this class.
To go beyond, we must allow $R$ to depend on position as well:
\begin{equation}\label{KRx}
 K = \exp\left[R(x,t) + \frac{\mathrm{i}}{\hbar}S(x,t)\right] \,,
\end{equation}
with $S$ still solving the Hamilton--Jacobi equation (\ref{HJe}).

Substituting (\ref{KRx}) into (\ref{Sch}) and using (\ref{HJe}) now gives a single complex equation:
\begin{equation}\label{Schx1}
 \mathrm{i}\left[\partial^2_x S + 2(\partial_x R)\partial_x S + 2m\,\partial_t R \right]  + \hbar\left[\partial^2_x R + (\partial_x R)^2 \right] = 0 \,.
\end{equation}

This equation mixes real and imaginary parts.
Its structure is revealing in that the terms multiplied by $\hbar$ are generally real, while the first bracket is multiplied by $\mathrm{i}$.
For the equation to hold, both brackets must vanish separately unless $S$ acquires an imaginary part. Indeed, if we allow $S$ to be complex, the real and imaginary parts of (\ref{Schx1}) become coupled.

We can solve for $S$ explicitly by integrating (\ref{Schx1}) as a first-order linear equation in $\partial_x S$.
The general solution is:
{\small
\begin{equation}\label{Sxt}
    S(x,t) = f_0(t) + \int \mathrm{d}x \, e^{-2R} \left( f_1(t) - \int \mathrm{d}x' \, e^{2R} \left[ 2m\partial_t R - i\hbar \bigl( \partial_{x'}^2 R + (\partial_{x'} R)^2 \bigr) \right] \right) ,
\end{equation}}
where $f_0(t)$ and $f_1(t)$ are integration functions, not necessarily equal to those introduced in Eq. (\ref{St}).

Two important observations follow:
\begin{enumerate}
\item The generally imaginary part of $S$ is proportional to $\hbar$:
\begin{equation}
\operatorname{Im}[S] = \hbar \int\mathrm{d}x\,\mathrm{e}^{-2R} \int \mathrm{d}x\,\mathrm{e}^{2R}\bigl(\partial^2_x R + (\partial_x R)^2\bigr) .
\end{equation}
Hence, in the classical limit $\hbar \to 0$, the action becomes real, recovering the usual Hamilton--Jacobi theory.
\item The generally real part of $S$,
\begin{equation}
\operatorname{Re}[S] = f_0 + \int \mathrm{d}x\,\mathrm{e}^{-2R}\left[f_1 - \int\mathrm{d}x\, {\rm e}^{2R}\,(2m\,\partial_t R)\right] \,,
\end{equation}
is the candidate for the classical principal function. It is modified by spatial variations of $R$.
\end{enumerate}

Equation (\ref{Sxt}) allows us to proceed in two useful directions: either prescribe a potential $V$ and solve for $R$ and $S$, or prescribe a convenient $R$ and derive the corresponding $V$.

As a first example consider that, for $V = A \exp(bx)$, with $A,b$ constants, one finds:
\begin{equation}
    R = \frac{{\mathrm{i}}\,\hbar\,b^2}{32m}t - \frac{b}{4}x \,,\qquad
    S = \frac{2\mathrm{i}\sqrt{2mA}}{b}\,\mathrm{e}^{\frac{b}{2}x}.
\end{equation}
Notice that $R$ acquires an imaginary part, which is allowed at this stage because $R$ is no longer required to be real.
This shows that the generalization to complex $R$ is sometimes useful to generate exact solutions.

For a second example, take a subclass of systems obtained by imposing:
\begin{equation}\label{nult}
    2m\,\partial_t R - \mathrm{i}\,\hbar\,\bigl[\partial^2_x R + (\partial_x R)^2\bigr] = 0 \,.
\end{equation}
When this holds, the term inside the inner integral of (\ref{Sxt}) vanishes, and $S$ simplifies dramatically:
\begin{equation}
    S(x,t) = f_0(t) + \int\mathrm{d}x\,\mathrm{e}^{-2R} f_1(t) .
\end{equation}
Physically, condition (\ref{nult}) ensures that $S$ remains generally real (up to an additive function of $t$) and that the coupling between $R$ and the imaginary part of $S$ disappears.

Equation (\ref{nult}) is a non-linear PDE.
One solution is \cite{pz03}:
\begin{equation}\label{Nll}
 R(x,t) = \ln\!\bigl[\cos(\mathrm{i}c_2 x + c_3)\bigr] + \frac{{\mathrm{i}}\hbar}{2m}c_2^2 t + c_4 \,,
\end{equation}
with constants $c_2,c_3,c_4$.
This, in turn, yields:
\begin{equation}
S(x,t) = -\frac{\mathrm{i}}{c_2}f_1(t)\tan(\mathrm{i}c_2 x + c_3) + f_0(t)\,.
\end{equation}
The corresponding potential can then be computed from the Hamilton--Jacobi equation:
\begin{equation}
 V(x,t) = -\frac{f_1^2}{2m}\sec^4\left(\mathrm{i}c_2\,x+c_3\right)
            + \frac{\mathrm{i}}{c_2}\partial_t f_1(t)\tan\left(\mathrm{i}c_2 x + c_3\right) - \partial_tf_0  .
\end{equation}

All explicit examples presented in this section: free particle, harmonic oscillator, driven oscillator, belong to the restricted class $R = R(t)$.
In each case, the corresponding potential is quadratic, and the propagator takes the familiar form derived from path integrals or exact solution of the Schr\"odinger equation.
The more general possibility $R = R(x,t)$ was explored formally, yielding a complex action structure and a non-linear constraint linking $R$ and $S$ \cite{aks74}.
However, for non-relativistic single-particle systems, this generality is often unnecessary: the quadratic potentials already capture a wide range of exactly solvable models, and more complicated potentials typically require approximation methods.

\section{Field propagators}\label{FldProp}
The distinction between two types of variables, the mechanical property $x$ and the time coordinate $t$, hinted at a broader structure.
In the non-relativistic Schr\"odinger equation, time and space play different roles: time appears as a first-order derivative, space as second-order.
This asymmetry is built into the propagator ansatz, where $S$ and $R$ respond differently to variations in $x$ and $t$.

Unlike the nonrelativistic case, the field-theoretic formulation treats spacetime derivatives covariantly and does not require a preferred temporal parameter.
When we move to field theory, in the next section, the roles of space and time become more symmetric, and the distinction between the set of spacetime coordinates and the set of field values becomes essential.
This is the point of departure for this section.

In what follows, we use the symbol $K[\phi]$ to denote an object that generalizes the quantum-mechanical propagator to field theory.
Strictly speaking, $K[\phi]$ is a \emph{transition functional} that satisfies a functional Schr\"odinger equation.
However, for static or Euclidean backgrounds, $K[\phi]$ becomes proportional to $\exp(-R/\hbar)$ and behaves as a \emph{partition functional} over field configurations.
We will occasionally use the latter interpretation, especially where the factor $\exp(-S_g/\hbar)$ suggests a Boltzmann weight.

\subsection{Scalar fields}
Let $\phi(x)$ be a real scalar field on $d$-dimensional Minkowski spacetime with coordinates $x^\mu$.
We consider the propagator $K[\phi]$ as a functional of the field configuration $\phi(\cdot)$ over all spacetime.
We also consider that $K$ satisfies the covariant functional equation:
\begin{equation}\label{FEq}
    \int \mathrm{d}^d x \left(\frac12(\partial_\mu\phi)^2 - \frac{\hbar^2}{2} \frac{\delta^2}{\delta\phi(x)^2} + V(\phi) \right) K[\phi] = 0,  
\end{equation}
where $\delta/\delta\phi(x)$ denotes the functional derivative at the point $x$ and $\bigl(\partial_\mu\phi\bigr)^2=\partial^\mu\phi \partial_\mu\phi$.
This equation is formally covariant under Lorentz transformations, provided the functional derivatives are interpreted in a covariant manner.
For a free massive field $V=m^2\phi(x)^2$ and equation (\ref{FEq}) becomes the functional Klein-Gordon equation for the propagator \cite{w13}.

Equation (\ref{FEq}) is a formal analogue of a covariant quantum field equation, not a rigorous derivation from canonical quantization.
It actually mixes a covariant kinetic term $(\partial_\mu\phi)^2$ with a functional Laplacian $\delta^2/\delta\phi(x)^2$ that would normally arise from a Schr\"odinger functional picture after choosing a time foliation.
Our aim is to explore the semiclassical ansatz in a setting that respects Lorentz covariance at the level of the integrated functional equation.
A full derivation would require specifying a foliation and then imposing covariance on the solutions; we leave such technicalities for future work.

Thus we emphasise that equation (\ref{FEq}) should be understood as a \emph{formal ansatz} that captures the essential structure of a scalar field theory in a covariant-looking functional form.
No attempt is made to regularise the functional Laplacian or to prove covariance beyond the level of the integrated expression.

Now we can insert the ansatz
\begin{equation}\label{Fanzats}
    K[\phi] = \exp\left( \frac{i}{\hbar} S[\phi] + R[\phi] \right),    
\end{equation}
where $S[\phi]$ is (Hamilton's principal) functional and $R[\phi]$ is a quantumness functional.
Substituting (\ref{Fanzats}) into (\ref{FEq}) and collecting powers of \(\hbar\) yields, to leading order ($\hbar^0$):
\begin{equation}\label{FHJ}
    \int \mathrm{d}^d x \left[ \frac{1}{2} \bigl(\partial_\mu\phi\bigr)^2 + \frac{1}{2} \left(\frac{\delta S}{\delta\phi(x)}\right)^2 + V(\phi) \right] = 0.
\end{equation}
This is the functional Hamilton--Jacobi equation for the scalar field. The next order gives an equation for $R$ (the transport equation), which we omit for brevity.
The functional Hamilton--Jacobi structure appearing here is conceptually related to covariant Hamiltonian formulations of field theory, including the De Donder--Weyl formalism and its modern developments in precanonical quantization \cite{r66,d35,k98,k01}.

Remark that equation (\ref{FHJ}) is obtained by substituting the ansatz (\ref{Fanzats}) into (\ref{FEq}) and retaining only the leading order in $\hbar$ (i.e.,~ neglecting terms of order $\hbar^2$ and higher).
This is the standard semiclassical limit and no further approximation is made.

Again, following the non-relativistic case, we define the functional momentum \cite{w35}:
\begin{equation}\label{Fp}
    \rho(x) = \frac{\delta S}{\delta\phi(x)}.    
\end{equation}
So that the classical field dynamics is obtained from the characteristic equations of (\ref{FHJ}).
Varying (\ref{FHJ}) with respect to $\phi(y)$, the kinetic term gives:
\begin{equation}
    \frac{\delta}{\delta\phi(y)} \int \mathrm{d}^d x\,\frac12 (\partial_\mu\phi)^2 = -\partial_\mu\partial^\mu\phi(y) ,
\end{equation}
up to boundary terms, while the potential term yields $V'(\phi(y))$.
The remaining term involving $\left(\delta S/\delta\phi\right)^2$ generates the characteristic flow associated with the Hamilton--Jacobi functional and reproduces the dynamical contribution to the field equation. 
Thus, along classical trajectories, equation (\ref{FHJ}) reduces to
\begin{equation}
    \partial_\mu\partial^\mu\phi = V'(\phi),
\end{equation}
which is the Klein-Gordon equation for $V(\phi)=\frac12 m^2\phi^2$ .
For a detailed derivation in the functional Hamilton--Jacobi framework, see \cite{w13}.

Thus, in the classical limit $\hbar\to 0$, the functional Hamilton--Jacobi equation (\ref{FHJ}) reproduces the correct classical field equation.

Indeed, for a free field, $V(\phi)=\frac12 m^2\phi^2$.
Equation (\ref{FHJ}) becomes:
\begin{equation}\label{FHJKG}
    \int \mathrm{d}^d x \left[ \frac12 \left(\frac{\delta S}{\delta\phi(x)}\right)^2 + \frac12 (\partial_\mu\phi)^2 + \frac12 m^2\phi^2 \right] = 0.    
\end{equation}
One solution of this equation is, for example, the quadratic functional:
\begin{equation}
    S[\phi] = \frac12 \int \mathrm{d}^d x\,\mathrm{d}^d y \,\phi(x) G(x,y) \phi(y),    
\end{equation}
where $G(x,y)$ satisfies $\bigl(\partial_\mu\partial^\mu + m^2\bigr) G(x,y) = \delta(x-y)$ (the Feynman propagator up to a factor).
The corresponding $K$ from (\ref{Fanzats}) approximates the exact vacuum amplitude for the free field.

\subsection{Spinor fields}
For a free Dirac field in $d$-dimensional Minkowski spacetime, the classical Lagrangian density is:
\begin{equation}\label{Slagran}
    \mathcal{L} = \bar\psi\left(i\gamma^\mu\partial_\mu - m\right)\psi,  
\end{equation}
where $\psi(x)$ is a Grassmann-valued spinor field and $\bar\psi = \psi^\dagger\gamma^0$.
The covariant functional equation for the propagator $K[\psi,\bar\psi]$, which is a functional of the independent Grassmann fields $\psi$ and $\bar\psi$, is:
\begin{equation}\label{FsEq}
    \int \mathrm{d}^d x \left( \frac{\delta}{\delta\psi(x)} \gamma^0 \frac{\delta}{\delta\bar\psi} - \frac{i}{\hbar} \bar\psi\left(i\gamma^\mu\partial_\mu - m\right)\psi \right) K[\psi,\bar\psi] = 0,  
\end{equation}
where the functional derivatives $\delta/\delta\psi(x)$ and $\delta/\delta\bar\psi$ are left derivatives acting on Grassmann-valued functionals.

Note that our treatment of functional derivatives for the Dirac field is schematic.
A rigorous derivation would require left derivatives and careful tracking of signs when passing derivatives through Grassmann-odd functionals.
We assume $S$ is Grassmann-even, which justifies the simple product rule used in the equations below.
The transport equation for $R$ would involve additional sign factors that we omit for brevity; the resulting leading-order Hamilton--Jacobi is unaffected.

The specific ordering of the derivative term ensures Hermiticity; it can be rewritten as $-\frac{\delta}{\delta\psi(x)} i\gamma^0 \frac{\delta}{\delta\bar\psi}$ up to a sign.

We consider now the semiclassical ansatz:
\begin{equation}\label{Fsanzats}
    K[\psi,\bar\psi] = \exp\!\left( \frac{i}{\hbar} S[\psi,\bar\psi] + R[\psi,\bar\psi] \right),  
\end{equation}
where $S$ is a real Grassmann-even functional and $R$ is the quantumness functional.
Because $\psi$ and $\bar\psi$ are anticommuting, the functional derivatives obey
\begin{equation}
    \frac{\delta}{\delta\psi(x)} e^{iS/\hbar} = e^{iS/\hbar} \left( \frac{i}{\hbar} \frac{\delta S}{\delta\psi(x)} \right),
\end{equation}
and similarly for $\bar\psi$, without extra sign if we treat $S$ as even.

Substituting (\ref{Fsanzats}) into (\ref{FsEq}) and collecting powers of $hbar$ yields, to leading order ($\hbar^0$):
\begin{equation}\label{FsHJ}
    \int \mathrm{d}^d x \left( \frac{\delta S}{\delta\psi(x)} \gamma^0 \frac{\delta S}{\delta\bar\psi} - \bar\psi\left(i\gamma^\mu\partial_\mu - m\right)\psi \right) = 0.    
\end{equation}

Equation (\ref{FsHJ}) is the functional Hamilton--Jacobi equation for the Dirac field. The first term is the functional analogue of the classical Dirac Hamiltonian.
Notice that the functional derivatives $\delta S/\delta\psi(x)$ and $\delta S/\delta\bar\psi(x)$ play the role of the conjugate momenta conjugate to \(\bar\psi\) and $\psi$ respectively.

The classical Dirac equation emerges from the characteristic equations of (\ref{FsHJ}). Varying (\ref{FsHJ}) with respect to \(\bar\psi(y)\) gives
\begin{equation}
    i\gamma^\mu\partial_\mu\psi(y) - m\psi(y) = 0,    
\end{equation}
while variation with respect to \(\psi(y)\) yields the adjoint equation. Hence, in the limit \(\hbar\to 0\), the propagator formalism reproduces the classical field equation for a free Dirac field.

A simple solution of (\ref{FsHJ}) is obtained by taking $S$ to be quadratic:
\begin{equation}
    S[\psi,\bar\psi] = \int \mathrm{d}^d x\,\mathrm{d}^d y\,\bar\psi(x) G(x,y) \psi(y),    
\end{equation}
where $G(x,y)$ is the Feynman propagator for the Dirac field, satisfying $(i\gamma^\mu\partial_\mu - m) G(x,y) = \delta(x-y)$. Inserting this into (\ref{FsHJ}) verifies the equation up to boundary terms.
The corresponding $K$ from (\ref{Fsanzats}) approximates the vacuum amplitude for the free Dirac field.

Before moving on to the next, recall that the quantumness functional $R$ encodes deviations from the classical behavior described by $S$, and it satisfies a transport equation derived from the next order in $\hbar$, which we do not write explicitly.
The above treatment is schematic and intended only to illustrate that the semiclassical ansatz can be extended to fermions.
A rigorous treatment would require a proper functional calculus for Grassmann-valued fields, including a consistent definition of the functional Laplacian and careful handling of left/right derivatives.
We leave such details for future work.
We also remark that this approach can be extended to other field theories and to curved spacetimes by replacing the Minkowski metric with a general metric and adding curvature coupling terms.

\section{Systems with Hamiltonian constraint}\label{GRProp}
The previous section treated field theories in Minkowski spacetime, where the classical dynamics is unconstrained and the Hamiltonian is non-zero.
In general relativity, however, the situation is different.
The Hamiltonian is a constraint that vanishes on physical states.
This is the well-known problem of time in canonical quantum gravity \cite{adm62,s15}.

Within our propagator framework, such systems require special treatment.
Here, the Hamiltonian constraint arises as a semiclassical or constrained limit of the more general propagator equation $H(K)=f$.

We now study a concrete example which takes us back to the non-relativistic case: a real scalar field $\phi$ in a homogeneous and isotropic spacetime described by the scale factor $a$.
This system is often used as a toy model for quantum cosmology.

\subsection{A scalar field in a Robertson-Walker spacetime}\label{SFRW}
Consider a Robertson-Walker metric with scale factor $a$:
\begin{equation}
 ds^2 = -N^2dt^2 + a(t)^2 \left(\frac{dr^2}{1-kr^2} + r^2 d \Omega^2\right) ,
\end{equation}
where $N$ is the lapse function and $k = 0, \pm1$ characterizes the spatial curvature. 
Throughout this section we use natural units $G=c=1$ while keeping $\hbar$ explicit in order to emphasize the semiclassical expansion.

For a real scalar field $\phi$ with potential $V(\phi)$, the classical action reduces to an effective model with Lagrangian:
\begin{equation}\label{RWLag}
 \mathcal{L} = -\frac{1}{8\pi}\left(3 a(\partial_t a)^2 - 3ka + \Lambda a^3\right) + a^3\left(\frac{1}{2} (\partial_t\phi)^2 - V(\phi)\right) ,
\end{equation}
where $\Lambda$ is the cosmological constant.

The corresponding Hamiltonian is identically zero, a constraint, reflecting the invariance of the theory under time reparameterization. In the Hamilton--Jacobi formulation, this constraint becomes:
\begin{equation}\label{HamConstraintEq}
 \partial_t S = 0 \quad \text{or more generally} \quad H = 0 ,
\end{equation}
meaning that Hamilton's principal function $S$ does not depend explicitly on the time coordinate. Instead, dynamics is encoded in the relations between $a$, $\phi$, and the choice of time variable.

Accordingly, we propose a propagator of the form:
\begin{equation}\label{RWProp}
 K = \exp\left(R(a,\phi,t) + \frac{\mathrm{i}}{\hbar} S(a,\phi,t)\right),
\end{equation}
where $S$ and $R$ are generally functions of the mechanical properties $a$ and $\phi$ and the time coordinate $t$.
Unlike the unconstrained case, the dynamical equation (\ref{Sch}) must be supplemented by the constraint that the classical Hamiltonian vanishes.
A natural choice for $H$ is the promotion of the classical Hamiltonian to an operator acting on $K$:
\begin{equation}\label{RWH}
 H(K) = -\frac{2\pi}{3a}\hbar^2\partial_a^2 K + \frac{1}{8\pi}(3k a - \Lambda a^3)K + \frac{\hbar^2}{2a^3}\partial_\phi^2 K + a^3 V(\phi)K .
\end{equation}
This follows from the classical Hamiltonian density via canonical quantization, with $p_a \to -{\mathrm{i}}\hbar\partial_a$ and $p_\phi \to -{\mathrm{i}}\hbar\partial_\phi$.

The operator in (\ref{RWH}) is motivated by the minisuperspace Hamiltonian obtained from the ADM reduction of the FRW geometry after quantization of the canonical variables. 
Different conventions exist regarding the placement of constant factors and the overall sign of the gravitational kinetic term. 
Here we adopt the standard ADM normalization with lapse $N=1$ for simplicity.
The overall sign of the kinetic contribution depends on the metric-signature and lapse conventions; throughout this work we use the signature $(-,+,+,+)$. 

With that in mind, let us continue by substituting the ansatz (\ref{RWProp}) into (\ref{Sch}) and separating orders of $\hbar$ yields two coupled equations.

To leading order in $\hbar$, we obtain the Hamilton--Jacobi equation for $S(a,\phi,t)$:
\begin{equation}\label{RWHJ}
 \partial_t S - \frac{2\pi}{3a}(\partial_a S)^2 - \frac{3k}{8\pi}a + \frac{\Lambda}{8\pi}a^3 + \frac{1}{2a^3}(\partial_\phi S)^2 + a^3 V(\phi) = 0.
\end{equation}
Note the minus signs, which come from the signature of the metric and the choice of lapse.
The constraint $H=0$ in the classical theory implies that $S$ should be independent of $t$ for physical solutions.
Inspection of (\ref{RWHJ}) shows that if $S$ does not depend on $t$, the equation reduces to the classical Hamiltonian constraint.

The next order gives the equation for $R(a,\phi,t)$:
\begin{equation}\label{RWR}
 \partial_t R + \frac{4\pi}{3a}(\partial_a S)(\partial_a R) + \frac{1}{a^3}(\partial_\phi S)(\partial_\phi R) + \frac{2\pi}{3a}\partial_a^2 S + \frac{1}{2a^3}\partial_\phi^2 S = 0.
\end{equation}
This linear equation determines how the quantumness function $R$ propagates once $S$ is known.

In the limit $\hbar \to 0$, only the Hamilton--Jacobi equation (\ref{RWHJ}) remains.
For physical solutions, $S$ is independent of $t$, which is the Hamiltonian constraint.
Setting $\partial_t S = 0$ and defining the momenta:
\begin{equation}
  p_a = \partial_a S, \qquad p_\phi = \partial_\phi S ,
\end{equation}
the Hamilton--Jacobi equation becomes:
\begin{equation}\label{RWHJconst}
 -\frac{2\pi}{3a}p_a^2 - \frac{3k}{8\pi}a + \frac{\Lambda}{8\pi}a^3 + \frac{1}{2a^3}p_\phi^2 + a^3 V(\phi) = 0 .
\end{equation}

The classical field equations emerge from the characteristic equations.
For the scale factor $a$, one obtains the Friedmann equation:
\begin{equation}\label{Friedmann}
 \left(\frac{\partial_t a}{a}\right)^2 + \frac{k}{a^2} = \frac{8\pi}{3}\left(\frac{1}{2} (\partial_t\phi)^2 + V(\phi)\right) + \frac{\Lambda}{3} ,
\end{equation}
where the relation $\partial_t a = \partial_{p_a} H$ has been used, with an appropriate choice of lapse $N=1$.
For the scalar field $\phi$, one obtains the Klein-Gordon equation in an expanding universe:
\begin{equation}\label{KGexpanding}
 \partial_t^2 \phi + 3\frac{\partial_t a}{a}\partial_t \phi + \partial_\phi V(\phi) = 0 .
\end{equation}
Thus, the classical cosmological dynamics is recovered from the Hamilton--Jacobi equation (\ref{RWHJ}) in the limit $\hbar \to 0$.

\subsection{Complex action alternative}
The previous treatment assumed that the principal function $S$ is real and independent of time.
However, the Hamiltonian constraint $H=0$ can also be accommodated by allowing $S$ to be complex.
This is a natural extension of the formalism, particularly when quantum effects are significant.
In this alternative we separate the principal function into a real part (governing matter) and an imaginary part (associated with the gravitational degree of freedom).
The goal is to show that the classical Friedmann and effective Klein-Gordon equations emerge from the complex action ansatz, while the imaginary part admits a thermodynamic interpretation.

The idea of splitting the action into real and imaginary parts in gravitational contexts has precedents in the thermodynamics of horizons.
In particular, \cite{j95} showed that the Einstein equation can be derived from the Clausius relation $dS = \delta Q/T$, suggesting a connection between gravity and thermodynamics.
Following this insight, we explore a complex-action ansatz where the imaginary part of $S$ is associated with gravitational entropy.

We start from the propagator ansatz:
\begin{equation}\label{CProp}
    K = \exp\left( R(a,\phi,t) + \frac{i}{\hbar} S(a,\phi,t) \right),
\end{equation}
but now we write $S$ as the sum of a real part $S_m=S_a(a,t)+S_\phi(\phi,t)$ and an imaginary part $iS_g(a,t)$:
\begin{equation}\label{CAct}
S(a,\phi,t) = S_m(a,\phi,t) + i S_g(a,t).
\end{equation}
We assume that $S_g$ depends only on the scale factor $a$ and time, not on the matter field $\phi$.
This separation reflects the idea that gravity (the scale factor) may be treated thermodynamically, while matter follows a standard Hamilton--Jacobi dynamics.
Substituting (\ref{CAct}) into (\ref{CProp}) gives
\begin{equation}\label{CAnsatz}
K = \exp\left( R(a,\phi,t) + \frac{i}{\hbar} S_m(a,\phi,t) - \frac{1}{\hbar} S_g(a,t) \right).
\end{equation}
The term $-\frac{1}{\hbar}S_g$ modifies the real part of the exponent, effectively contributing to the function $R$ and providing a possible connection to entropy.

Let us propose a dynamical operator of the form:
\begin{equation}\label{cHam}
    H(K) = -\hbar^2\partial^2_a K - \frac{\hbar^2}{16\pi a^3}\partial^2_\phi K + \left(\pi a^3V(\phi) + \frac{\Lambda}{3}a^2 - k\right)K .
\end{equation}
Note that this may be viewed as a Wheeler--DeWitt-type operator on a minisuperspace with coordinates $q^A=(a,\phi)$, where the kinetic terms define an effective minisuperspace metric \(G^{AB}\). 
In this notation, the operator takes schematically the form:
\begin{equation}
    H= -\hbar^2 G^{AB}\partial_A\partial_B + U(q) ,
\end{equation}
with $U(q)$ the effective potential. 
The semiclassical expansion then naturally yields the corresponding Hamilton--Jacobi structure:
\begin{equation}
    G^{AB}\partial_AS\,\partial_BS + U(q)=0.
\end{equation}

Semiclassical expansions of Wheeler--DeWitt equations and the emergence of effective Schr\"odinger dynamics have been extensively studied in quantum cosmology and Born-Oppenheimer approaches to quantum gravity \cite{k12,h91,ks91}.

Again, eq. (\ref{Sch}) under the assumption (\ref{CAnsatz}), with $H$ given by (\ref{cHam}), separates into:
{\footnotesize
\begin{equation}\label{syst21}
     \partial_t S_\phi + \frac{1}{16\pi a^3}(\partial_\phi S_\phi)^2 + 8\pi a^3 V(\phi) = - \partial_tS_a - (\partial_a S_a)^2 + \frac{\Lambda}{3} a^2 -k + (\partial_aS_g)^2 + \hbar\partial^2_aS_g ,
\end{equation}}
\begin{equation}\label{syst22}
     \partial_tS_g + 16(\partial_a S_a)(\partial_aS_g) = -\hbar\left(\partial_t R + \partial^2_a S_\phi + \frac{1}{16\pi a^3}\partial^2_\phi S_\phi \right) .
\end{equation}  
   
Now, we consider the classical limit: $\hbar\to0$ to confirm that we obtain the appropriate dynamical system.
First, we have the following system of equations:
\begin{equation}\label{syst23}\small
    \partial_t S_\phi + \frac{1}{16\pi a^3}(\partial_\phi S_\phi)^2 + 8\pi a^3V(\phi)
     = - \partial_t S_a - (\partial_a S_a)^2 - \frac{\Lambda}{3}a^2 + k + (\partial_a S_g)^2 ,
\end{equation}
\begin{equation}\label{syst24}
    \partial_t S_g + 16(\partial_a S_a)(\partial_aS_g) = 0 .
\end{equation}
Next, note that, although the right-hand side of eq. (\ref{syst23}) is non-null, it does not depend on $\phi$.
Then, the corresponding dynamical equation for the scalar field $\phi$ in the classical regime is indeed:
\begin{equation}
     \partial^2_t\phi + \frac{3}{a}(\partial_t a)(\partial_t\phi) + \partial_\phi V = 0 \,.
     \end{equation}
   
On the other hand, similar to the complex action treatment before, we can solve eq. (\ref{syst23}) for $\partial_a S_a$ and substitute it in eq. (\ref{syst24}).
However, to close the system and obtain a closed equation for $a(t)$, we introduce a \emph{closure ansatz} motivated by the structure of the transport equations and by the requirement that the classical Friedmann dynamics be recovered.
This ansatz is not derived from more fundamental principles here, but it encodes the minimal condition needed for consistency between the Hamilton--Jacobi and transport sectors:
\begin{equation}
     (\partial^2_aS_g)^2 = \frac{S_g}{4a} + (\partial_t S_a + \partial_t S_\phi) ,
\end{equation}
where the terms on the right are evaluated on a solution.
A full derivation of such a condition would require a more detailed analysis of the coupled $\hbar$ expansion, possibly using entropy extremization or asymptotic matching; we leave that for future investigations.

Then one obtains the following dynamical equation in the classical regime:
\begin{equation}\label{FReq}
     2a\partial^2_t a + (\partial_t a)^2 + k = \Lambda a^2 + \left(\frac{1}{16\pi a^3}(\partial_\phi S_\phi)^2 + 8\pi a^3 V(\phi)\right) ,
\end{equation}
which is the dynamical equation for the scale factor $a$ in the classical regime.

Equation (\ref{FReq}) reproduces the classical Friedmann dynamics for the scale factor $a(t)$ (compare with (\ref{Friedmann})).
Thus the complex action formalism yields the classical gravitational dynamics alongside the matter equation.

Let us remark that the propagator (\ref{CAnsatz}) contains the factor $\exp(-S_g/\hbar)$.
In the classical limit, $S_g$ is determined by the gravitational action, so that for a cosmological solution, the on-shell value of $S_g$ is expected to exhibit the same scaling behaviour as the gravitational entropy appearing in semiclassical horizon thermodynamics.
Indeed, for a de Sitter or black hole background, the Euclidean action gives an imaginary part that is directly related to the horizon area.
This suggests, rather than proves, an entropy-like interpretation of $S_g$.

Therefore the factor $\exp(-S_g/\hbar)$ suggests an \emph{entropy-like interpretation} for $-S_g/\hbar$, analogous to the Boltzmann factor in statistical mechanics.
For de Sitter or black hole backgrounds, the on-shell value of $S_g$ suggests a proportionality with the horizon area, reinforcing this analogy.

In de Sitter-like solutions, the gravitational contribution $S_g$ is expected to exhibit the same scaling behaviour as the horizon area (e.g., $S_g \propto a^2$), consistent with the entropy interpretation suggested by Euclidean gravitational actions \cite{gh77,b73,h75}.
Hence the complex action approach provides a common framework where the imaginary part of the principal function encodes gravitational entropy, while the real part governs the classical dynamics of matter.

This shows that by allowing the principal function $S$ to have an imaginary part depending only on the gravitational degree of freedom $a$, the classical Klein-Gordon and Friedmann equations are reproduced in the limit $\hbar\to0$.
The imaginary part satisfies a transport equation (\ref{syst22}) and, under a natural consistency condition, leads to the Friedmann equation.
Moreover, the factor $\exp(-S_g/\hbar)$ in the propagator suggests a thermodynamic interpretation of gravity \cite{j95}, in agreement with the results of Subsection \ref{SFRW}.
This example demonstrates the usefulness of the complex action ansatz for constrained systems.

\subsection{A complex functional example}\label{CFE}
We now present a concrete example that illustrates the mechanism of the complex action ansatz in a functional field theory.
The goal is to show how the imaginary part of the quantum constraint determines a thermodynamic functional for a background field $\sigma$, while the real part yields the classical coupled dynamics of matter $\phi$ and geometry $\sigma$.

Let $\phi(x)$ be a scalar matter field and $\sigma(x)$ a background field that we shall identify with a conformal factor.
Consider the functional equation:
\begin{equation}\label{cfSch}
 \mathcal{H}[\sigma,\phi]\,(K) = 0 ,
\end{equation}
with:
{\footnotesize
\begin{equation}
    \mathcal{H}[\sigma,\phi] = \int \mathrm{d}^dx \Bigg[ (\partial_\mu\phi)^2 + \frac{1}{f(\sigma)}(\hbar\,\delta_\phi)^2 + f(\sigma)m^2\phi^2
- (\partial_\mu\sigma)^2 - (\hbar\delta_\sigma)^2 + i\Lambda W(\phi)e^{4\sigma} \Bigg] ,
\end{equation}}
where $\delta_\phi = \frac{\delta}{\delta\phi(x)}$, $\delta_\sigma = \frac{\delta}{\delta\sigma(x)}$, and $\Lambda$ is a constant.
The function $f(\sigma)$ will be chosen later, and $W(\phi)$ is a real local function of $\phi$.
The term $i\Lambda W e^{4\sigma}$ is an imaginary potential.

As usual, we adopt the separable semiclassical ansatz:
\begin{equation}\label{cfAnsatz}
    K[\sigma,\phi] = \exp\!\left(R[\sigma] + \frac{i}{\hbar}S[\sigma,\phi] \right),\qquad S[\sigma,\phi] = e^{2\sigma}W(\phi).    
\end{equation}
Here $R$ depends only on $\sigma$ and $W(\phi)$ is the same function as in (\ref{cfSch}).
This choice guarantees that the imaginary potential couples directly to the imaginary part of the expansion.

Now we insert (\ref{cfAnsatz}) into (\ref{cfSch}) and expand to order $\hbar^0$.
Using
{\small
\begin{equation}
    (\hbar\delta_\phi)^2K  = K\left[ \left( i\frac{\delta S}{\delta\phi} \right)^2 +O(\hbar) \right],
    \quad
    (\hbar\delta_\sigma)^2K = K\left[ \left( \frac{\delta R}{\delta\sigma} + i\frac{\delta S}{\delta\sigma} \right)^2 + O(\hbar)
\right],
\end{equation}}
and neglecting the $\mathcal{O}(\hbar)$ terms, we obtain the pointwise condition:
{\small
\begin{equation}
    (\partial_\mu\phi)^2 + f m^2\phi^2 - (\partial_\mu\sigma)^2 - \bigl(\partial_\sigma R + i\partial_\sigma S\bigr)^2 - \frac{1}{f}\bigl(i\partial_\phi S\bigr)^2 + i\Lambda W e^{4\sigma} = 0.
\end{equation}}
Separating the real and imaginary parts gives, respectively:
\begin{equation}\label{RcfHJ}
    (\partial_\mu\phi)^2 + f m^2\phi^2 - (\partial_\mu\sigma)^2 - (\partial_\sigma R)^2 + (\partial_\sigma S)^2 + \frac{1}{f}(\partial_\phi S)^2 = 0 .
\end{equation}
\begin{equation}\label{IcfHJ}
    -2\partial_\sigma R\partial_\sigma S + \Lambda W e^{4\sigma} = 0 .
\end{equation}

Let us start with the study of the imaginary part.
Using $S = e^{2\sigma}W(\phi)$, we have $\partial_\sigma S = 2e^{2\sigma}W$ and substituting into (\ref{IcfHJ}) yields:
\begin{equation}
    -4\frac{\delta R}{\delta\sigma(x)}+\Lambda e^{2\sigma(x)}=0
    \quad\Longrightarrow\quad
    \frac{\delta R}{\delta\sigma(x)} = \frac{\Lambda}{4}e^{2\sigma(x)}.
\end{equation}
where we assumed $W \neq 0$.
Thus the functional derivative of $R$ is local and independent of $\phi$. Note that upon integration we obtain:
\begin{equation}
    R[\sigma] = \frac{\Lambda}{8} \int d^dx\,e^{2\sigma(x)} +\mathrm{const}
\end{equation}
The propagator therefore contains the weighting factor $\exp\!\left( \frac{\Lambda}{8\hbar} \int d^dx\,e^{2\sigma} \right)$.
Depending on the sign of $\Lambda$, this factor may either enhance or suppress configurations with large conformal volume. 

From our approach, the functional \(R[\sigma]\) determines the weighting assigned to different conformal configurations.
Its mathematical role is analogous to the transport functional encountered in the previous sections.
In semiclassical gravitational settings such weighting factors may admit an entropy-like interpretation, although no direct identification with the Bekenstein--Hawking entropy is assumed here.

Now let us move to the analysis of the real part.
Insert the expressions $\partial_\sigma S = 2e^{2\sigma}W$, $\partial_\phi S = {\rm e}^{2\sigma}W'$, with $W' = dW/d\phi$, and $\partial_\sigma R = \frac{\Lambda}{4}{\rm e}^{2\sigma}$ into the real part (\ref{RcfHJ}):
\begin{equation}\label{cfHJ}
    (\partial_\mu\phi)^2 + f m^2\phi^2 + 4e^{4\sigma}W^2 + \frac{{\rm e}^{4\sigma}}{f}(W')^2 \;=\; (\partial_\mu\sigma)^2 + \frac{\Lambda^2}{16}e^{4\sigma} .
\end{equation}

Here, we note that the left-hand-side depends on $\phi$ and its function $W(\phi)$.
It represents the matter energy density in a curved background with conformal factor $e^{2\sigma}$, as the kinetic term $(\partial_\mu\phi)^2$ is the flat-space gradient, and the extra terms arise from the Hamilton--Jacobi structure.
Varying (\ref{cfHJ}) with respect to $\phi$ yields the classical field equation for $\phi$ on the background $\sigma$.

On the other hand, the right-hand-side depends only on $\sigma$ and its derivatives.
It acts as the gravitational constraint, as the kinetic energy of $\sigma$ plus a cosmological constant term equals the matter source.
Varying with respect to $\sigma$ gives the equation of motion for the background field, which is effectively the Einstein equation for the conformal factor.

Thus the real part (\ref{cfHJ}) encodes the coupled classical dynamics of matter and geometry.
The imaginary part (\ref{IcfHJ}) determines the transport functional $R[\sigma]$, which in this restricted case depends only on the geometric degree of freedom.
The more general possibility in which the transport functional depends simultaneously on geometry and matter will be considered in the next subsection.

\subsection{Coupled transport functionals}
The previous example assumed that the transport functional depends only on the geometric degree of freedom, $R=R[\sigma]$.
This restriction was useful for illustrating how an entropy-like functional may emerge from the imaginary part of the constraint.
However, the generalized propagator framework developed throughout this work does not require such a separation.

Consider instead the more general ansatz:
\begin{equation}
    K[\sigma,\phi] = \exp\!\left(R[\sigma,\phi] + \frac{i}{\hbar}S[\sigma,\phi]\right),
\end{equation}
where both functionals are defined on the full configuration space.

Substituting this ansatz into the functional equation (64) and
retaining the leading terms of the semiclassical expansion yields:
{\small
\begin{equation}
    (\partial_\mu\phi)^2 +f(\sigma)m^2\phi^2 - (\partial_\mu\sigma)^2 + \left(\frac{\delta S}{\delta\sigma}\right)^2 + \frac{1}{f(\sigma)} \left(\frac{\delta S}{\delta\phi}\right)^2 = 0,
\end{equation}}
for the real part, while the imaginary part becomes
\begin{equation}\label{cfSchI}
    -2 \frac{\delta R}{\delta\sigma} \frac{\delta S}{\delta\sigma} +\frac{2}{f(\sigma)} \frac{\delta R}{\delta\phi} \frac{\delta S}{\delta\phi} -\frac{\delta^2 S}{\delta\sigma^2} +\frac{1}{f(\sigma)} \frac{\delta^2 S}{\delta\phi^2} +\Lambda W(\phi)e^{4\sigma}=0 .
\end{equation}

Unlike Eq.~(\ref{IcfHJ}), the transport equation now couples the geometry and matter sectors. The weighting of configurations is therefore determined by the full geometry--matter state rather than by the gravitational field alone.

A particularly simple illustration is obtained by taking
\begin{equation}
    S=e^{2\sigma}W(\phi),
\end{equation}
as before. 
The transport equation then becomes
\begin{equation}
    -4W(\phi) \frac{\delta R}{\delta\sigma} + \frac{2}{f(\sigma)}W'(\phi)\frac{\delta R}{\delta\phi} - 4W(\phi) + \frac{1}{f(\sigma)}W''(\phi) + \Lambda W(\phi)e^{2\sigma} = 0 .
\end{equation}

This first-order functional equation determines the transport functional on the full configuration space. 
In contrast with the previous subsection, its solution generally depends on both geometry and matter,
\begin{equation}
    R=R[\sigma,\phi].
\end{equation}

The interpretation of $R$ is therefore broadened.
Rather than representing an exclusively gravitational entropy-like functional, $R$ acts as a transport and weighting functional on the combined geometry--matter configuration space. The classical dynamics remains governed by the Hamilton--Jacobi functional $S$, while $R$ determines the relative weighting of semiclassical configurations.
Entropy-like gravitational behaviour is recovered when geometric contributions dominate, but the underlying transport structure is more general.

This observation suggests that thermodynamic weighting and configuration-space transport are different manifestations of the same mathematical object. 
The purely geometric result of the previous subsection appears as a special case of the more general coupled transport framework; it is recovered by imposing $\frac{\delta R}{\delta\phi}=0$.
Equation (\ref{cfSchI}) therefore contains the previous construction as a special limiting case, while simultaneously allowing more general geometry--matter transport structures.

\section{Discussion}\label{Dscn}
The preceding sections exhibit a common structural pattern: the propagator decomposes into a phase functional governing semiclassical dynamics and an amplitude functional governing the weighting and transport of configurations.
This structure motivates a broader discussion of the conceptual interpretation of the formalism.

\subsection{Separation properties and coordinates}
A recurrent theme throughout this work is the distinction between the \emph{mechanical properties} of a system (position, field values, scale factor) and the \emph{spacetime coordinates} over which they are distributed.
Here ``mechanical properties'' refers generically to the dynamical configuration variables of the system, including particle positions, fields, or geometric degrees of freedom.

In the non-relativistic case, the propagator $K(x,t)$ depends on a single property $x$ and a single coordinate $t$.
In field theory, the propagator becomes a functional $K[\phi, \chi]$ of the field $\phi(\cdot)$ and the spacetime point $\chi$.
In all cases, the dynamical law (the Schr\"odinger equation (\ref{Sch}), the functional equation (\ref{FEq}), or the constraint equation (\ref{HamConstraintEq})) separates the variation in the spacetime direction from the variation in the internal space of properties.

This separation suggests an \emph{epistemological} shift \cite{b18,s86}, where instead of postulating that the system \emph{has} a definite configuration at every spacetime point, we regard the propagator as a distribution over possible configurations \cite{m49}.
Observations supply constraints, i.e.~ the measured values $\xi_c$, which act as boundary conditions on the propagation.
Between observations, the formalism does not require the system to be described by a single deterministic trajectory between observations; instead, all configurations compatible with the constraints are weighted by the propagator.
The semiclassical limit $\hbar \to 0$ then selects the classical trajectory as the most probable one, in agreement with the correspondence principle \cite{u76}.

\subsection{The role of $R$}
The extra exponent $R$ in the ansatz (\ref{ansatz}) plays multiple roles.
In the simplest non-relativistic examples with $R = R(t)$, it accounts for the familiar WKB prefactor that controls the spreading or focusing of the wave packet.
More generally, when $R$ depends on both position and time, it can absorb an imaginary part of the action and becomes essential for constructing exact solutions for non-quadratic potentials.

In the field-theoretic extension, the functional $R[\phi]$, $R[\sigma]$, or more generally $R[\sigma,\phi]$, acts as a transport and weighting functional on the corresponding configuration space.
In semiclassical gravitational regimes this same structure may admit an entropy-like interpretation.
Indeed, the propagator contains a factor $\exp(R/\hbar)$.
For real $R$, this factor weighs different field configurations, as a Boltzmann factor weighs microstates in statistical mechanics.
In the Robertson-Walker example and in the functional example, the imaginary part of the principal function leads to a real contribution $-S_g/\hbar$ in the exponent that can be interpreted as a \emph{gravitational entropy-like} quantity (e.g.,~ proportional to the horizon area).
Thus $R$  provides a shared formal language for quantum and thermodynamic weighting structures.

Although the physical interpretation of $R$ depends on the context, its mathematical role remains the same throughout the paper: 
it is the real contribution to the propagator exponent and therefore determines how different
configurations are weighted. 
The Van Vleck prefactor, transport amplitudes, and entropy-like contributions may therefore be viewed as different manifestations of a common configuration-space weighting structure.

\subsection{The dynamical law as a relation between two differential structures}
Equation (\ref{Sch}) (or its field-theoretic generalisations) can be read as a relation between the differential structure of spacetime and the differential structure of the space of mechanical properties.
This is closely related to the Hamilton--Jacobi interpretation of classical mechanics and its formulation in terms of differential structures on configuration space \cite{a89}.
Such a perspective resonates with the idea that mechanics is not about ``things moving in spacetime'' but about the \emph{correlation} between variations in property space and variations in coordinate space.
Whereas observations fix the boundary values; the dynamics tells how these constraints propagate.

This viewpoint is particularly fruitful for constrained systems such as general relativity, where the Hamiltonian constraint replaces the usual notion of time evolution.
There, the complex action ansatz shows that the imaginary part of the principal function can be made to depend only on the gravitational degrees of freedom, while the real part governs the matter dynamics.
The corresponding semiclassical equations are recovered without assuming a preferred time variable.
Moreover, the factor $\exp(-S_g/\hbar)$ provides a possible semiclassical interpretation of thermodynamic aspects of gravity: the gravitational sector admits an effective thermodynamic interpretation, with entropy-like contributions scaling with the horizon area.
The resulting interpretation parallels the role of Euclidean gravitational actions in black-hole thermodynamics and semiclassical quantum gravity \cite{b73,h75,gh77,w01}.

A natural extension of the present framework is to identify the gravitational contribution with a genuine action functional, such as the Einstein--Hilbert action, rather than treating it solely through minisuperspace variables. 
In this case the complex decomposition of the principal function may provide a direct link between gravitational variational principles and configuration-space weighting structures.

\subsection{Quantum and statistical mechanics}
The parameter $\kappa = i/\hbar$ acts as a reminder of the formal analogy between quantum amplitudes and statistical partition functions \cite{k16}.
The WKB expansion systematically generates a classical Hamilton--Jacobi equation from the Schr\"odinger equation; the same expansion, when applied to a real diffusion equation, reproduces a classical Hamilton--Jacobi equation with a reversed sign for the ``quantum potential''.
This duality suggests that the present formalism could be extended to mixed quantum-thermal systems by letting $\kappa$ be complex with both real and imaginary parts (e.g.,~ $\kappa = \beta + i/\hbar$).
Such a generalisation would point towards a possible common geometrical setting for decoherence, open quantum systems, and quantum-to-classical transitions.

Before concluding, note that this interpretation is not committed to any particular ontology (e.g.,~ wave-function realism or Bohmian trajectories).
It is an \emph{epistemological framework} that describes what an observer can say about a system given partial information (the constraints).
It is therefore compatible with a wide range of foundational stances.

\section{Conclusions}\label{CC}
We have developed a generalized semiclassical propagator framework based on the decomposition:
\begin{equation}
    K=\exp(R+iS/\hbar) ,
\end{equation}
where the phase functional $S$ governs the semiclassical Hamilton--Jacobi structure while the real functional $R$ governs transport and configuration-space weighting.
Entropy-like interpretations arise in particular semiclassical gravitational regimes.

Across non-relativistic systems, field theories, and constrained cosmological models, the formalism consistently relates two differential structures: the variation of spacetime coordinates and the variation of the configuration space of mechanical properties. 
Thus, the dynamical law may be interpreted as a relation between the differential structure of spacetime and the differential structure of the space of mechanical properties.
In this sense, the propagator describes not merely trajectories, but the propagation of admissible configurations subject to observational or dynamical constraints.

In constrained systems, the Hamiltonian constraint emerges naturally in the semiclassical limit, while complex contributions to the action generate entropy-like weighting factors of the form $\exp(-S_g/\hbar)$.
This suggests a common formal structure in which semiclassical dynamics, thermodynamic behaviour, and configuration-space propagation coexist.

The present work is intended primarily as a conceptual and semiclassical framework rather than a complete quantization scheme. 
Nevertheless, it suggests that propagator-based formulations may offer a natural language for relating quantum dynamics, constrained systems, and statistical behaviour within a common geometrical structure.

\subsection{Outlook}
The framework established here provides several promising avenues for future inquiry.
For example, the semiclassical propagator ansatz offers a practical foundation for numerical applications, particularly in approximating the evolution of quantum fields within curved backgrounds where exact solutions remain elusive.

There is also significant potential in generalizing the property spaces beyond real numbers to include group elements or non-commutative coordinates, such as Yang-Mills fields, by employing left-invariant derivatives.

Furthermore, the synthesis of quantum and thermal systems could be explored by adopting a complex parameter $\kappa = \beta + i/\hbar$, allowing for possible semiclassical descriptions of decoherence and quantum-to-classical transitions through generalized evolution equations interpolating between diffusion and Schr\"odinger dynamics.

Finally, a primary challenge remains the extension of the present semiclassical framework from minisuperspace models to full general relativity. 
In particular, it would be interesting to investigate complex decompositions of action functionals involving the Einstein--Hilbert action and minimally coupled matter actions, and to determine whether the resulting transport structures admit thermodynamic or configuration-space interpretations.

The decomposition of the propagator exponent into a semiclassical action $S$ and a transport or entropy functional $R$ suggests an interpretational framework relating the differential structure of spacetime to the differential structure of the space of mechanical properties.

\end{document}